# Effect of the adsorption component of the disjoining pressure on foam film drainage

Stoyan I. Karakashev[1], Anh V. Nguyen[2] and Roumen Tsekov[1]
[1]Department of Physical Chemistry, University of Sofia, 1164 Sofia, Bulgaria
[2]Department of Chemical Engineering, University of Queensland, Brisbane, QLD 4072, Australia

The present work is trying to explain a discrepancy between experimental observations of the drainage of foam films from aqueous solutions of sodium dodecyl sulfate (SDS) and the theoretical DLVO-accomplished Reynolds model. It is shown that, due to overlap of the film adsorption layers, an adsorption component of the disjoining pressure is important. The pre-exponential factor of the adsorption component was obtained by fitting the experimental drainage curves. It corresponds to a slight repulsion, which reduces not only the thinning velocity as observed experimentally but corrects also the film equilibrium thickness.

Karakashev et al.[1] have investigated the drainage of foam films of dilute aqueous solution of sodium dodecyl sulfate within the concentrations range 1-100 μM. They tried to describe the kinetics of foam film thinning by the Reynolds lubrication approximation accounting for the Marangoni effect, surface shear viscosity and DLVO forces. Significant discrepancy between the theoretical prediction and the experimental results was observed. The detailed analysis showed that the deviation of the theory from the experimental data originates from the interaction between the film surfaces. Therefore, it was concluded that the classical DLVO theory only is not sufficient to match the experimental data. It was suggested that the discrepancy between theory and experiment is due to a neglected variation of the adsorption component of the surface tension during the film drainage. Large number of literature confirms the applicability of the DLVO theory to foam films. However, number of papers[2-6] report deviations of this theory from experimental data. This discrepancy is pronounced mostly in thin films between hydrophobic surfaces. To solve the problem some authors[7-11] introduced in the theory an additional non-DLVO force, the so-called hydrophobic force, which can be attractive or repulsive.[12-14] There is number of attempts in the literature[15-19] to explain the nature of the hydrophobic interaction but still no full agreement of the opinions is reached. The classical DLVO theory does not account also for other interactions in the thin liquid films. For instance, the interactions between the overlapping diffusive adsorption layers should contribute to the overall interaction between the film surfaces and this contribution should increase with decreasing film thickness.

The idea of the adsorption interaction between the film surfaces originates from the work of Ash, Everett and Radke[20] and it is further developed by the Russian school of colloid chemistry. The dispersion interaction between the solutes and the film surfaces is accounted for [21-23] and it results in a correction in the van der Walls component of the disjoining pressure. This additional adsorption term in the total interaction between the film surfaces could be important but it has been overlooked in a large volume of literature causing diversity of the opinions regarding the hydrophobic interaction. The reason for this is that the researchers cited above have described the surfactant distribution only as a result of interactions with the surfaces but neglected the interactions between solute molecules. Of course, the latter are not important in dilute solutions far away from the surface, but when the adsorption is considered the concentration near a surface is tremendously increased. Tsekov and Schulze[17] suggested first a clear thermodynamic interpretation of the adsorption term in the total disjoining pressure. They called it hydrophobic force, since the origin of the adsorption is the surface hydrophobicity and the surfactant ability to reduce it. The aim of this paper is to employ this approach for explanation of our experimental data.[24] The good agreement will certainly draw attention on the importance of the adsorption disjoining pressure.

According to the thin liquid film thermodynamics any change of the film free energy $F$ at constant temperature is given by

$$dF = -pdV + \gamma dA + \sum \tilde{\mu}_i dn_i \qquad (1)$$

where the extensive film parameters are volume $V$, film area $A$ and number of moles $\{n_i\}$ of the film components. The relationship between the intensive parameters pressure $p$, film tension $\gamma$ and electrochemical potentials $\{\tilde{\mu}_i\}$ is given by the Gibbs-Duhem equation

$$-Vdp + Ad\gamma + \sum n_i d\tilde{\mu}_i = 0 \qquad (2)$$

It is known that the thin liquid films are anisotropic structures[25] and their pressure tensor possesses two distinct components, the normal and tangential ones. At equilibrium the normal component of the pressure tensor equals to the gas pressure outside, while the tangential component equals to the pressure in the meniscus adjacent to the film. The pressure $p$ is the normal component of the pressure tensor. The film tension $\gamma$ consists in two additives,[26] where $h = V/A$ is the film thickness,

$$\gamma = 2\sigma + \Pi h \qquad (3)$$

The purely interfacial part is twice the film surface tension $\sigma$ while the 'bulk' part is accounted by the disjoining pressure $\Pi$. Introducing Eq. (3) in Eq. (1) the latter changes to

$$dF = -pAdh - (p-\Pi)hdA + 2\sigma dA + \sum \tilde{\mu}_i dn_i \qquad (4)$$

It is obvious now that the normal and tangential components of the film pressure tensor are not equal and the disjoining pressure is their difference.

Using Eq. (3) one can derive an alternative form of Eq. (2)

$$-Vdp + Ad(2\sigma + \Pi h) + \sum n_i d\tilde{\mu}_i = 0 \tag{5}$$

After Gibbs the film can be idealized by filling it with the bulk liquid from the meniscus. Hence, subtracting from Eq. (5) the Gibbs-Duhem relation $dp_L = \sum c_i d\tilde{\mu}_i$ for the liquid in the meniscus, where $\{c_i\}$ are the concentrations of the chemical components there, and keeping in mind that $\Pi = p - p_L$, one yields an important interfacial Gibbs-Duhem relation[27]

$$d\sigma = -\sum \Gamma_i d\tilde{\mu}_i - \Pi dh/2 \tag{6}$$

where $\{\Gamma_i = (n_i - c_i V)/2A\}$ are the component adsorptions. Eq. (6) provides straightforward an important definition of the disjoining pressure as the thickness derivative of the film surface tension

$$\Pi = -2(\frac{\partial \sigma}{\partial h})_{\tilde{\mu}} \tag{7}$$

as well as the following Maxwell relation for the disjoining pressure

$$(\frac{\partial \Pi}{\partial \tilde{\mu}_i})_h = 2(\frac{\partial \Gamma_i}{\partial h})_{\tilde{\mu}} \tag{8}$$

The latter already hints the important effect of adsorption on the disjoining pressure.[28,29]

Since the surfactants could be charged species the film surface tension depends on electrostatics as well. It can be split into superposition of water, electrostatic and adsorption components, $\sigma = \sigma_W + \sigma_{EL} + \sigma_{AD}$, which are independent if the surface potential $\phi_s$ does not depend on the film thickness. Thus, during the film drainage the adsorptions and surface charge density, respectively, can vary but the electrostatic component $\sigma_{EL}$ will not be affected by. At constant temperature the water component depends only on the film thickness, while the surfactant component depends on the adsorption. Substituting this presentation in Eq. (7) the disjoining pressure splits also into three distinct components

$$\Pi = \Pi_{VW} + \Pi_{EL} - 2\sum (\frac{\partial \sigma_{AD}}{\partial \Gamma_i})(\frac{\partial \Gamma_i}{\partial h})_{\tilde{\mu}} \qquad (9)$$

where $\Pi_{VW} = -2(\partial \sigma_W / \partial h)_{\tilde{\mu}}$ and $\Pi_{EL} = -2(\partial \sigma_{EL} / \partial h)_{\tilde{\mu}}$ are the well-known van der Waals and electrostatic components. Indeed, at low surface potentials the electrostatic component of the surface tension equals to $\sigma_{EL} = -\varepsilon_0 \varepsilon \kappa \phi_s^2 \tanh(\kappa h/2)/2$, where $\kappa$ is the reciprocal Debye length, and the corresponding electrostatic disjoining pressure $\Pi_{EL} = \varepsilon_0 \varepsilon \kappa^2 \phi_s^2 / 2\cosh^2(\kappa h/2)$ acquires its classical form.[30]

Let us consider now the last adsorption component of the disjoining pressure in Eq. (9). To calculate its thickness dependence of adsorption one can employ the Maxwell relation (8). Introducing the following definition $\Delta_h X \equiv X(h) - X(\infty)$ for a difference between the values of a property $X$ of the equilibrium films with thickness $h$ and infinity, respectively, one can write $\Pi = -\Delta_h p_L$. Note, that changing the film thickness only its tangential pressure component changes, while the normal one $p_G = p_L(h=\infty)$ remains constant. Thus, the Maxwell relation (8) can be consecutively modified to

$$2(\frac{\partial \Gamma_i}{\partial h})_{\tilde{\mu}} = (\frac{\partial \Pi}{\partial \tilde{\mu}_i})_h = -\Delta_h (\frac{\partial p_L}{\partial \tilde{\mu}_i})_h = -\Delta_h c_i \qquad (10)$$

Knowing the adsorption isotherm $c_i(\Gamma_i)$ at constant surface potential one is able to integrate this equation to obtain the thickness dependence of adsorption. If the changes of the concentration and adsorption, respectively, are small one can employ the following linear relationship $\Delta_h \Gamma_i \approx a_i \Delta_h c_i$, where $a_i = (\partial \Gamma_i / \partial c_i)_{h=\infty}$ is the adsorption length on a single flat liquid/gas interface. The latter, representing the thickness of the adsorption layer, depends on the adsorption equilibrium constants and $\phi_s$. Solving now the linearized differential equation (10) yields

$$\Delta_h \Gamma_i = \Delta_0 \Gamma_i \exp(-\frac{h}{2a_i}) \qquad (11)$$

where $\{\Delta_0 \Gamma_i\}$ is the difference between the adsorptions in a surfactant bilayer ($h \to 0$) and on a single flat surface ($h \to \infty$). Substituting now this expression into the definition of the adsorption disjoining pressure from Eq. (9) leads to

$$\Pi_{AD} = \sum (\frac{\partial \sigma_{AD}}{\partial \Gamma_i}) \frac{\Delta_0 \Gamma_i}{a_i} \exp(-\frac{h}{2a_i}) \qquad (12)$$

Note that depending on the sign of $\{\Delta_0\Gamma_i\}$, the adsorption disjoining pressure can be either positive or negative. It could be also zero if no changes in the adsorption in a bilayer and on a single flat surface take place. This is probably the most widespread case, which explains why the adsorption disjoining pressure is still not well studied. The present thermodynamic theory cannot give any value of $\{\Delta_0\Gamma_i\}$ but just assuming them describes the thickness dependence of their effect.

The drainage of thin liquid films depends substantially on the mobility of film surfaces.[25] Our present estimates show, however, that within the specified SDS concentration range the Marangoni effect is always strong enough to block the tangential flow on the film interfaces. Hence, the drainage velocity can be well approximated by the classical Stefan-Reynolds equation

$$-\frac{dh}{dt} = \frac{2h^3(p_\sigma - \Pi)}{3\eta R^2} \qquad (13)$$

where $p_\sigma$ is the capillary pressure, $R$ is the film radius, and $\eta$ is the liquid viscosity.

The disjoining pressure in Eq. (13) is of crucial importance for the modeling of the drainage. How it was shown above, $\Pi$ is a superposition on the van der Waals, electrostatic and adsorption components. To determine the effect of the adsorption disjoining pressure correctly reliable expressions for the DLVO components are required. The van der Waals disjoining pressure between the film surfaces can be estimated from the expression $\Pi_{VW} = -A/6\pi h^3$. Since the film thickness $h$ is always larger than 150 nm and the Hamaker constant is about $A \approx 2\times10^{-21}$ J, the van der Waals disjoining pressure is negligible for the present system. At constant surface potential, the electrostatic disjoining pressure, calculated by the exact numerical solution of the non-linear Poisson-Boltzmann equation, is semi-analytically described as[31]

$$\Pi_{EL} = 32RTc\tanh^2(y/4)[(1+\cosh\kappa h)^{-1} + f\sinh^2(y/4)\exp(-f\kappa h)] \qquad (14)$$

where $f = 2\cosh(0.332|y| - 0.779)$ for $|y| \le 7$, $y = F\phi_s/RT$ is a dimensionless surface potential on a single flat air/solution interface. In the case of a single surfactant Eq. (12) reduces to

$$\Pi_{AD} = \Delta_0\sigma_{AD}\exp(-h/2a)/a \qquad (15)$$

where $\Delta_0\sigma_{AD}$ is the difference of the adsorption components of the surface tension on the bilayer and on a single flat interface. Here $a$ is the surfactant adsorption length. Since $\Delta_0\sigma_{AD}$ could be either positive or negative depending on the interactions between the two monolayers of the bilayer, $\Pi_{AD}$ could be also repulsive or attractive, respectively. In order to compare the above theory with the experimental data, Eq. (13) was numerically integrated using a fourth-step Runge-Kutta method. A macro was written in the VBA (Visual Basic for Application) programming language available from Microsoft Excel. The measured values for the zeta potential were adopted for the surface potential $\phi_s$ in $\Pi_{EL}$. The adsorption lengths $a$ are determined by the model of Kralchevsky et al.[32] The computed film thickness vs. time was compared with the experimental one for each of the SDS concentrations in order to obtain the best fit of the free parameter $\Delta_0 \sigma_{AD}$ (see Fig. 1).

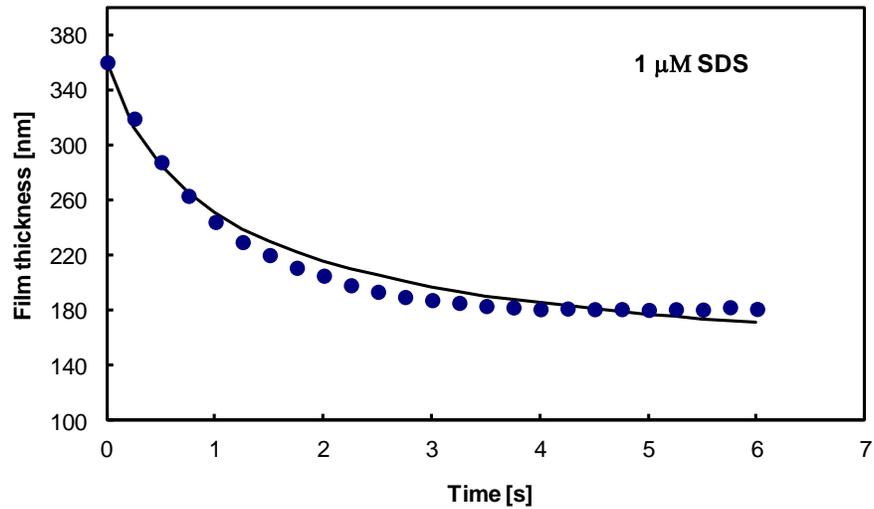

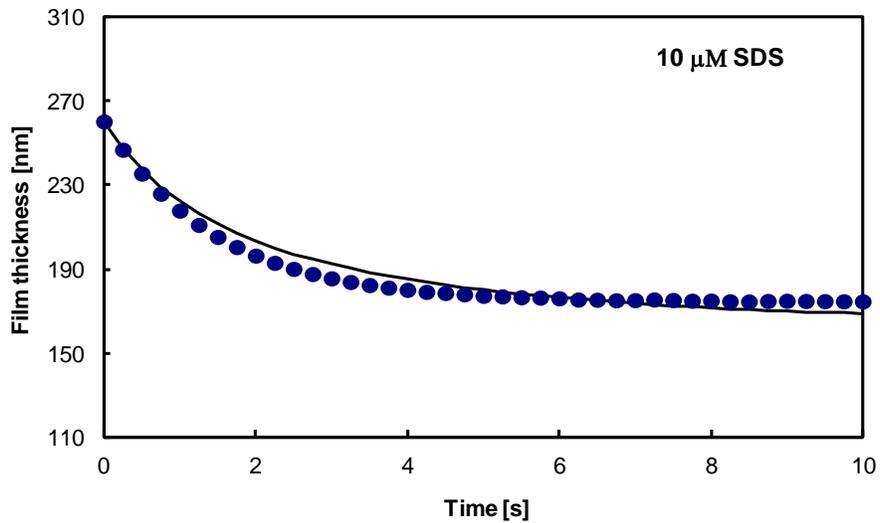

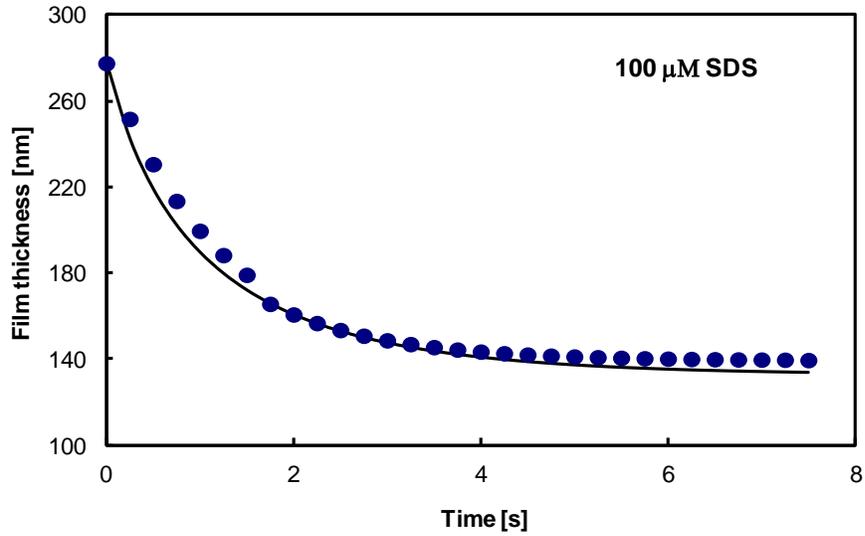

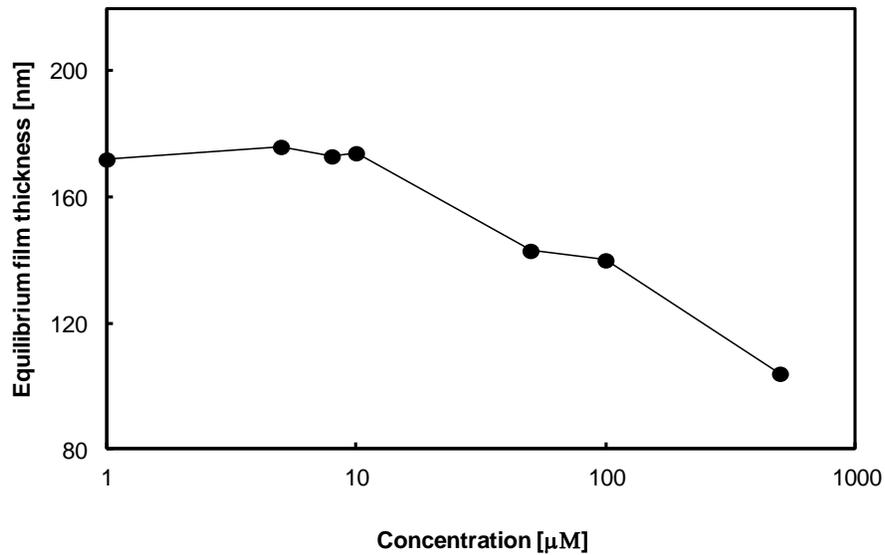

**Figure 1.** Examples of experimental data and theoretical fits of the films thickness vs. time for some SDS concentrations. The last plot presents the equilibrium film thickness vs. SDS concentration.

These data along with the capillary pressure are presented in Table 1. The juxtaposition between the experimental and theoretical film thickness vs. time is presented in the complementary material attached to the paper. Table 1 shows that the adsorption disjoining pressure is positive since $\Delta_0 \sigma_{AD} > 0$, which corresponds to additional (non-DLVO) repulsion between the film surfaces. This repulsion originates from the overlap of the adsorption layers of the two film surfaces. The latter indicates that the adsorption component of the film surface tension increases with the decrease of the film thickness. Hence, the surfactant adsorption for this particular case (SDS) diminishes during the film thinning. In general, the adsorption disjoining pressure should disappear at zero SDS concentration. As expected $\Delta_0\sigma_{AD}$ increases with increasing surfactant concentration. Since the adsorption length $a$ reduces with increasing of $c$, the adsorption disjoining pressure becomes shorter ranged and stronger.

Table 1. Capillary pressure $p_\sigma$, zeta potential $\phi_s$, adsorption length $a$ and fitting parameter $\Delta_0\sigma_{AD}$ vs. concentration of SDS

| $c$ [μM] | $p_\sigma$ [Pa] | $\phi_s$ [mV] | $a$ [nm] | $\Delta_0\sigma_{AD}$ [μN/m] |
|---|---|---|---|---|
| 1 | 72.6 | -63.0 | 275 | 7 |
| 5 | 72.5 | -56.1 | 275 | 7 |
| 8 | 72.5 | -52.7 | 275 | 7 |
| 10 | 72.4 | -49.2 | 275 | 7 |
| 50 | 72.4 | -48.6 | 274 | 7 |
| 100 | 71.8 | -48.0 | 274 | 10 |
| 500 | 70.0 | -52.8 | 249 | 13 |

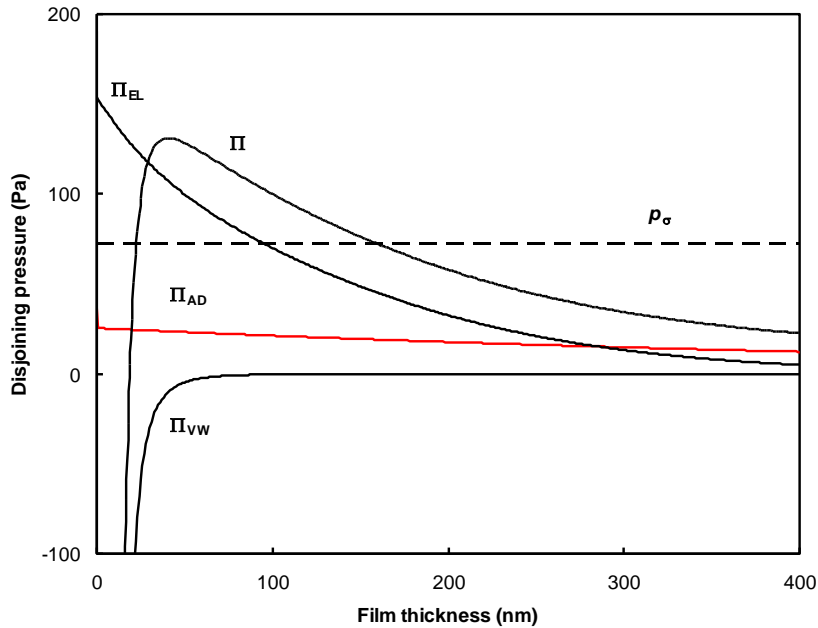

Figure 2. Electrostatic $\Pi_{EL}$, van der Waals $\Pi_{VW}$, adsorption $\Pi_{AD}$ and total $\Pi$ disjoining pressures vs. the film thickness for 10 μM SDS solution.

A general question here is how large is the contribution of the adsorption disjoining pressure into the total disjoining pressure. Another question is if this non-DLVO force is long or short ranged. One can find answers of these questions in Fig. 2, which shows the electrostatic $\Pi_{EL}$, van der Waals $\Pi_{VW}$, adsorption $\Pi_{AD}$ and total $\Pi$ disjoining pressures vs. the film thickness. The straight dashed line in the figure represents the sucking capillary pressure $p_\sigma$. As seen the adsorption interaction between the film surfaces is decaying upon the film thickness much weaker than the electrostatic and van der Waals interactions. The contribution of the adsorption interaction to the total interaction between the film surfaces is significantly small. However, above a given film thickness (ca. 300 nm for the particular case of 10 µM SDS) the adsorption interaction prevails over the electrostatic one. Despite being long-ranged, the adsorption disjoining pressure cannot become equal to the capillary pressure at any film thickness due to its small value. This means that in absence of electrostatic disjoining pressure an equilibrium film cannot be formed and the film will thin until rupture. The absence of adsorption interaction, however, will reflect in a significantly smaller equilibrium thickness of the film (95 nm instead of 160 nm) and faster film drainage.

The present paper proves the existence of adsorption non-DLVO disjoining pressure between the foam film surfaces. It originates by the overlap between the adsorption layers and can be attractive, repulsive or vanishing. The adsorption disjoining pressure is related to the properties of the adsorption layers. It is part of the hydrophobic interaction between the film surfaces.[17] If the surfactant adsorption diminishes upon the decrease of the film thickness the adsorption interaction is repulsive and vice versa. We mention here as well that such films develop streaming potential upon their drainage.[33] This theory is validated by experiment on kinetic of thinning of foam films from SDS within the concentration range 1-500 µM. Fit upon the parameter $\Delta_0 \sigma_{AD}$ for each one of the concentrations is performed. As expected $\Delta_0 \sigma_{AD}$ increases with increase on the surfactant concentration. Thus defined the adsorption interaction does not differ from this one defined by Tsekov and Schulze[17] and Wang and Yoon.[12,13] A more detailed study for the effect of the adsorption isotherm on the adsorption component of the disjoining pressure can be found in Ref.[34]

1. Karakashev, S.I.; Manev, E.D.; Nguyen, A.V., *Colloids Surf. A* 2008, **319**, 34
2. Exerowa, D.; Kolarov, T.; Khristov, K., *Colloids Surf.* 1987, **22**, 171
3. Buchavzov, N.; Stubenrauch, C., *Langmuir* 2007, **23**, 5315
4. Mishra, N.C.; Muruganathan, R.M.; Müller, H.-J.; Krustev, R., *Colloids Surf. A* 2005, **256**, 77
5. Bowen, W.R.; Doneva, T.A.; Stoton, J.A.G., *Colloids Surf. A* 2002, **201**, 73
6. Krustev, R.; Müller, H. J., *Langmuir* 1999, **15**, 2134
7. Israelachvili, J.N.; Pashley, R.M.; Perez, E.; Tandon, R.K., *Colloids Surf.* 1981, **2**, 287
8. Israelachvili, J.; Pashley, R., *Nature (London)* 1982, **300**, 341
9. Rabinovich, Y.I.; Derjaguin, B.V., *Colloids Surf.* 1988, **30**, 243
10. Fa, K.; Nguyen, A.V.; Miller, J.D., *J. Phys. Chem. B* 2005, **109**, 13112

11. Eriksson, J.C.; Ljunggren, S.; Claesson, P.M., *J. Chem. Soc., Faraday Trans. 2* 1989, **85**, 163
12. Wang, L.; Yoon, R.-H., *Langmuir* 2004, **20**, 11457
13. Wang, L.; Yoon, R.-H., *Colloids Surf. A* 2005, **263**, 267
14. Karakashev, S.I.; Phan, C.; Nguyen, A.V., *J. Colloid Interface Sci.* 2005, **291**, 489
15. Schalchli, A.; Sentenac, D.; Benattar, J.J.; Bergeron, V., *J. Chem. Soc., Faraday Trans.* 1996, **92**, 2317
16. Evans, D.R.; Craig, V.S.J.; Senden, T.J., *Physica A* 2004, **339**, 101
17. Tsekov, R.; Schulze, H.J., *Langmuir* 1997, **13**, 5674
18. Yoon, R.-H.; Aksoy, B.S., *J. Colloid Interface Sci.* 1999, **211**, 1
19. Basu, S.; Nandakumar, K.; Masliyah, J.H., *J. Colloid Interface Sci.* 1996, **182**, 82
20. Ash, S.G.; Everett, D.H.; Radke, C., *J. Chem. Soc., Faraday Trans.* 1973, **69**, 1256
21. Derjaguin, B.V.; Churaev, N.V., *Kolloidn. Zh.* 1975, **37**, 1075
22. Derjaguin, B.V., *Colloid Polym. Sci.* 1980, **258**, 433
23. Exerowa, D.; Churaev, N.V.; Kolarov, T.; Esipova, N.E.; Panchev, N.; Zorin, Z.M., *Adv. Colloid Interface Sci.* 2003, **104**, 1
24. Karakashev S.I.; Nguyen A.V., *Colloids Surf. A* 2007, **293**, 229
25. Ivanov, I.B. (Ed.), *Thin Liquid Films*, Marcel Dekker, New York, 1988
26. Rusanov, A.I., *Phase Equilibria and Surface Phenomena*, Khimiya, Leningrad, 1967
27. Toshev, B.V.; Ivanov, I.B., *Colloid Polym. Sci.* 1975, **253**, 558
28. Derjaguin, B.V.; Starov, V.M.; Churaev, N.V., *Colloid J. (USSR)* 1976, **38**, 411
29. Vassilieff, C.S.; Toshev, E.T.; Ivanov, I.B., in *Surface Forces and Boundary Liquid Layers*, B.V. Derjaguin (Ed.), Nauka, Moscow, 1983, p. 168
30. Verwey, E.J.W.; Overbeek, J.T.G., *Theory of the Stability of Lyophobic Colloids*, Elsevier, Amsterdam, 1948; p. 218
31. Nguyen, A.V.; Evans, G.M.; Jameson, G.J., *J. Colloid Interface Sci.* 2000, **230**, 205
32. Kralchevsky, P.A.; Danov, K.D.; Broze, G.; Mehreteab, A., *Langmuir* 1999, **15**, 2351
33. Karakashev, S.I.; Tsekov, R., *Langmuir* 2011, **27**, 265
34. Tsekov, R., *Ann. Univ. Sofia, Fac. Chem.* 2011, **102/103**, 273